 \documentclass[twocolumn]{aastex61}
\hypersetup{linkcolor=red,citecolor=green,filecolor=cyan,urlcolor=magenta}

\usepackage{natbib}

\newcommand\aastex{AAS\TeX}

\usepackage{gensymb}
\newcommand{\lpobj}{\object{LP261-75B~}}
\submitjournal{ApJ}
\shorttitle{\aastex\ Rotational Modulations in an L-dwarf Companion}
\shortauthors{Manjavacas et al.}
\begin{document}


\title{Cloud Atlas: Discovery of Rotational Spectral Modulations in a Low-mass, L-type Brown Dwarf Companion to a Star}

\correspondingauthor{Elena Manjavacas}
\email{elenamanjavacas@email.arizona.edu}

\author{Elena Manjavacas}
\affil{Department of Astronomy/Steward Observatory, The University of Arizona, 933 N. Cherry Avenue, Tucson, AZ 85721, USA}

\author{D\'aniel Apai}
\affiliation{Department of Astronomy/Steward Observatory, The University of Arizona, 933 N. Cherry Avenue, Tucson, AZ 85721, USA}
\affiliation{Department of Planetary Science/Lunar and Planetary Laboratory, The University of Arizona, 1640 E. University Boulevard, Tucson, AZ 85718, USA}
\affiliation{Earths in Other Solar Systems Team, NASA Nexus for Exoplanet System Science.}


\author{Yifan Zhou}
\affiliation{Department of Astronomy/Steward Observatory, The University of Arizona, 933 N. Cherry Avenue, Tucson, AZ 85721, USA}

\author{Theodora Karalidi}
\affiliation{Department of Astronomy/Steward Observatory, The University of Arizona, 933 N. Cherry Avenue, Tucson, AZ 85721, USA}
\affiliation{Department of Astronomy and Astrophysics, University of California, Santa Cruz, California, USA}

\author{Ben W. P. Lew}
\affiliation{Department of Planetary Science/Lunar and Planetary Laboratory, The University of Arizona, 1640 E. University Boulevard, Tucson, AZ 85718, USA}

\author{Glenn Schneider}
\affiliation{Department of Astronomy/Steward Observatory, The University of Arizona, 933 N. Cherry Avenue, Tucson, AZ 85721, USA}

\author{Nicolas Cowan}
\affiliation{Department of Earth \& Planetary Sciences, 3450 University St. Montreal, Quebec, Canada H3A 0E8}

\author{Stan Metchev}
\affiliation{The University of Western Ontario, Department of Physics and Astronomy, 1151 Richmond Avenue, London, ON N6A 3K7, Canada}

\author{Paulo A. Miles-P\'aez}
\affiliation{The University of Western Ontario, Department of Physics and Astronomy, 1151 Richmond Avenue, London, ON N6A 3K7, Canada}
\affiliation{Department of Astronomy/Steward Observatory, The University of Arizona, 933 N. Cherry Avenue, Tucson, AZ 85721, USA}

\author{Adam J. Burgasser}
\affiliation{Center for Astrophysics and Space Science, University of California San Diego, La Jolla, CA 92093, USA}

\author{Jacqueline Radigan}
\affiliation{Utah Valley University, 800 West University Parkway, Orem, UT 84058, USA}

\author{Luigi R. Bedin}
\affiliation{INAF – Osservatorio Astronomico di Padova, Vicolo dell'Osservatorio 5, I-35122 Padova, Italy}

\author{Patrick J. Lowrance}
\affiliation{IPAC-Spitzer, MC 314-6, California Institute of Technology, Pasadena, CA 91125, USA}

\author{Mark S. Marley}
\affiliation{NASA Ames Research Center, Mail Stop 245-3, Moffett Field, CA 94035, USA}



\begin{abstract}
Observations of rotational modulations  of brown dwarfs and giant exoplanets allow the characterization of condensate cloud properties. As of now rotational spectral modulations have only been seen in three L-type brown dwarfs. We report here the discovery of rotational spectral modulations in {LP261-75B}, an L6-type intermediate surface gravity companion to an M4.5 star. As a part of the \textit{Cloud Atlas} Treasury program we acquired time-resolved Wide Field Camera 3 grism spectroscopy (1.1--1.69~$\mu$m) of {LP261-75B}. We find gray spectral variations with the relative amplitude displaying only a weak wavelength dependence and no evidence for lower-amplitude modulations in the 1.4~$\mu$m water band than in the adjacent continuum. 
The likely rotational modulation period is {4.78$\pm$0.95~h}, although the rotational phase is not well sampled. The minimum relative amplitude in the white light curve {measured over the whole wavelength range} is 2.41$\pm$0.14\%. We report an unusual light curve with seemingly three peaks {approximately evenly distributed} in rotational phase. The spectral modulations {suggests} that the upper atmosphere cloud properties in {LP261-75B} are similar to two other mid-L dwarfs of typical infrared colors, but differ from that of the extremely red L-dwarf WISE0047.
\end{abstract}




\section{Introduction} \label{sec:intro}

{Brown dwarfs and giant {non-irradiated} exoplanets {at appropriately young age}, share similar ranges of temperatures and atmospheric abundances, and have similar spectra influenced by condensate clouds that are common in their atmospheres \citep[and references therein]{Reid2008,Cruz,Kirkpatrick2010,Faherty2013}.  Due to these similarities, comparative studies of brown dwarfs and giant exoplanets provide powerful insights into the atmospheric structures and processes common to these objects.}

{In addition, Solar System giant planets -- that can be studied at great detail -- provide cooler and lower-mass analogs to brown dwarfs that help us understand the dynamics, composition, and structures in brown dwarf and exoplanet atmospheres.} Giant planets in our Solar System show bands, hot spots, zones, jets and storms that vary over time. \cite{Ackerman_Marley2001} suggested that similar heterogeneous cloud patterns and structures may be inferred also in brown dwarfs, resulting in photometric variability driven by the cloud structures in brown dwarf atmospheres. 
Rotational modulations due to heterogeneous clouds have been observed in Jupiter's light curve: {\citet{Karalidi2015} measured $U$ and $R$-band rotational modulations, they compared them with simultaneous \textit{Hubble Space Telescope} (HST) images, showing that hot spots in the atmosphere of Jupiter are responsible for its troughs in the light curve, and the Great Red Spot is responsible for the peaks.}
In addition, \cite{Simon2016} carried out a similar study on Neptune using \textit{Kepler} data and disk-resolved images from the \textit{Keck} 10-m telescope, obtaining analogous results.

Putative photometric variability for brown dwarfs {\citep[and references therein]{Tinney_Tolley1999, Bailer_Jones_Mundt1999, Bailer_Jones2001, Gelino2002, Koen2003, Morales_Calderon2006}} had also been reported almost since their discovery. {Most likely, the first robust and confirmed variability detection due to heterogeneous cloud coverage in the atmosphere of a brown dwarf  is  from \citet{Artigau2009}, for  2MASS~J0136565+093347 (SIMP0136)}. {Multiple analyses show that photometric variability of brown dwarfs  older than 10~Myr and with spectral types later than L3 is most likely due to heterogeneous cloud coverage  \citep{Artigau2009, Radigan2012, Apai2013, Buenzli2015, Metchev2015}. These conclusions are supported by the lack of correlation between H$\alpha$ emission and variability. This strongly suggests that magnetically induced spots (starspots) are not responsible for photometric variability in most brown dwarfs \citep{Miles_Paez2017}. In addition, because dust grains coagulate rapidly in brown dwarf disks \citep{Apai2005,Pascucci2009} and the {typical lifetimes of optically thick dust disks around brown dwarfs is less than 10~Myr \citep{Carpenter2006}}, remnant dust disks also cannot be the primary source of variability.
{Indeed, \cite{Radigan2012} compared time-resolved near-infrared photometric data to \cite{Saumon_Marley2008} atmospheric models for cloudy atmospheres with different cloud thickness, finding that models could only reproduce the overall shape of the (time-averaged) spectrum and the ratio of photometric variability amplitudes for the L/T transition brown dwarf 2MASS~J21392676+0220226 (2M2139), with changes in {\em both} the surface temperature and cloud coverage.} They demonstrated that the photometric variability of 2M2139 is originated from patchy cloud coverage: either from holes and cloud thickness variations. \cite{Apai2013} presented time-resolved HST near-infrared spectra and studied photometric and spectral variations, showing that the variability in two L/T transition brown dwarfs (2M2139 and SIMP0136) is caused by correlated cloud thickness and temperature variations. In 2014 \cite{Buenzli2014} modeled near-infrared spectral variations observed in Luhman 16B with HST and confirmed that the variability in this L/T transition object is also explained by correlated cloud thickness and temperature variations.}

{Non-axisymmetric cloud structures} in a rotating ultracool atmosphere introduce distinct {signatures} in their disk-integrated light curves. An observed light {curve} is a geometrically weighted, {observable}-disk-integrated function of the surface brightness distribution of the target \citep[]{Cowan_Agol2008, Cowan2017}. As the source rotates, {the projected position of cloud structures} on the {observable hemisphere} changes, {altering its disk-integrated intensity}. By modeling the observed light curve we can place constraints on the properties of the {cloud structures}, such as their location on the disk (longitude, latitude), size and contrast to the background atmosphere  \citep[]{Knutson2007,Cowan2013,Apai2013,Karalidi2015}. 

To date, published light curves of various ultracool dwarfs show single or double peaks, whose shapes vary with time, sometimes even within a single rotation \citep{Artigau2009, Radigan2012, Apai2013, Guillon2013, Buenzli2015,Metchev2015,Karalidi2015,Yang2016,Lew2016}. These observations are interpreted as single or multiple cloud structures  in the atmosphere of brown dwarfs  that can evolve at short timescales. {Based on a comprehensive Spitzer Space Telescope photometric monitoring program \cite{Apai2017} reported dramatic and continuous lightcurve evolution in six brown dwarfs, all of which showed a quasi-periodic lightcurve. For three of the targets — all at the L/T transition — they found that two or three planetary-scale waves (k=1 and k=2 waves, where k is the wavenumber) provided good fits to the lightcurves, if the k=1 waves had similar, but slightly different periods. These slightly different peaks in the power spectrum could be attributed to differential rotation, although that explanation is not unique. These results argue for zonal atmospheric circulation and planetary-scale waves in L/T transition brown dwarfs.}

In this paper, we present, analyze and discuss the {unusual} light curve of {LP261-75B} (L6 brown dwarf companion to a M4.5 star) with hints of three peaks.  In Section \ref{LP261-75B} we describe the  characteristics of {LP261-75B}. In Section \ref{Observations} we describe the observations acquired and the data reduction performed. In Section \ref{systematics}, we present the analysis we performed to rule out the possibility that systematics are influencing our results. In Section \ref{origin} we analyze the possible causes that would explain the spectro-photometric variability of LP261-75B. In Section \ref{Results} we summarize the results obtained from the analysis of the spectroscopy and light curves of {LP261-75B}. In Section \ref{Discussion} we examine the possible causes that could explain the {unusual} light curve of  the object. Finally, we summarize our conclusions in Section \ref{Conclusions}.

\section{LP261-75B}\label{LP261-75B}

{2MASSW J09510549+3558021  was discovered by \citet{Kirkpatrick_2000}. 2MASSW J09510549+3558021 has a L6 spectral type in the optical \citep{Kirkpatrick_2000}, and a L4.5 spectral type in the near infrared \citep{Allers&Liu}. Its   {$J$, $H$ and $K$-band magnitudes of 17.23$\pm$0.21, 15.89$\pm$0.14 and 15.14$\pm$0.14, respectively.} \cite{Burgasser2005} discovered that 2MASSW J09510549+3558021 (LP261-75B)  is  a companion of the LP261-75A active M4.5 star, with a separation of 12$^{\prime\prime}$.} The system is  at a trigonometric distance of $31.6\pm1.3$~pc \citep{Liu_Dupuy_Allers2016}.

{\cite{Reid_Walkowicz2006}  estimated an age of {100--300~Myr} for LP621-75A based on its coronal activity levels, which are comparable to stars with similar spectral types in the Pleiades open cluster (of age 125$\pm$8~Myrs, \citealt{Stauffer1998}). \cite{Shkolnik2009} estimated an age between 40 and 300~Myr through the X-Ray activity of LP251-75A.  {Assuming} an average age for LP261-75AB of 100--200~Myrs, evolutionary models \citep{Burrows} provide an estimated mass for {LP261-75B} between  15-30~$M_{\rm Jup}$ \citep{Artigau2015}. }

{LP261-75B's spectral energy distribution is similar to those typical to field L6 brown dwarfs, with similarly prominent FeH features in the H-band, and alkali lines in the near-infrared spectra  \citep{Liu_Dupuy_Allers2016}. These spectral characteristics are consistent with those expected for a L6 brown dwarf with age at the upper end of the given age range for LP261-75B (40--300~Myr, \citealt{Stauffer1998})}

\section{Observations and Data Reduction}\label{Observations}

LP261-75B was observed in Cycle 23 of the HST program (P.I. D. Apai, GO14241) using the Wide Field Camera 3 (WFC3) in the near infrared channel, and its G141 grism \citep{MacKenty2010}. WFC3 with the G141 grism covers the wavelength range between $\sim$1.05 and 1.7~$\mu$m, with a spectral resolving power of 130 at 1.4 $\mu$m. WFC3/IR has a plate scale of 0.13 arcsec/pixel. {We acquired 6 consecutive orbits of  observations on Dec 21, 2016 UTC}. To obtain an accurate wavelength reference,  a direct image in each orbit was also taken in the F132N filter. We used a 256$\times$256 subarray mode to eliminate mid-orbit buffer dumps.

We performed the data reduction using the same method as in previous works published by our group \citep{Apai2013, Buenzli2014, Yang2015, Lew2016}. {The uncertainty level for our spectra after the data reduction is 0.1--0.3\% per spectral bin, {measured} using the reduced spectra in the range of 1.1--1.69~$\mu$m to avoid the noise at the edges of the spectra due to the drop in the instrument sensitivity}. These uncertainties are due to photon noise, errors in the sky subtraction, and the read-out noise.

\begin{figure}[htb!]
\centering
\includegraphics[width=0.45\textwidth]{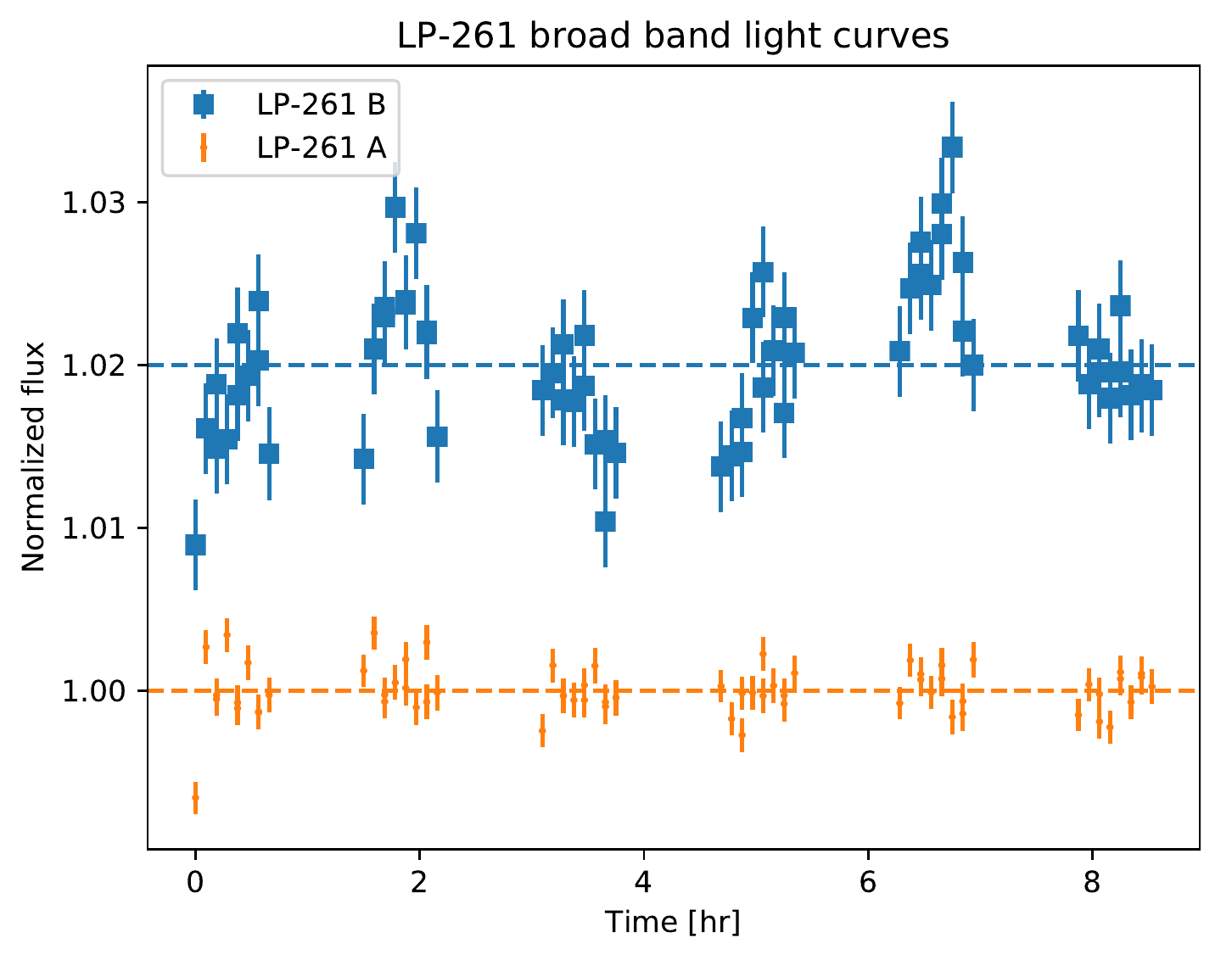}
\caption{\label{LC_AB} Broad band or white light curves (1.1--1.69$\mu$m) of LP261-75B (blue dots), and LP261-75A (orange dots). {The fluxes of the light curves  are normalized to their respective mean fluxes. We applied a positive shift in the y-axis to the LP261-75B light curve to overplot both light curves in the same plot}}.
\end{figure}

\section{Systematics Corrections}\label{systematics}

\subsection{Aperture selection}

{Due to the close angular proximity of the bright M4.5 star (LP261-75A) care must be taken when choosing the spectral extraction rectangular aperture, parallel to the direction. If the spectral extraction aperture is too wide, the spectra of \lpobj may be contaminated by the spectra of the primary; if the aperture  is  too narrow the signal-to-noise will be compromised. To {explore the effect of the aperture width choice}, we  have  performed  the data reduction using apertures ranging from {0.5} to  20$\times$FWHM (Full Width at Half Maximum), with steps of 0.5 FWHM (measured in pixels), at 1.4~$\mu$m centered at  the position of the target spectrum.  We measured the total flux  of the spectrum for each of those apertures. 
In the explored aperture width range we did not see an inflection in the flux that would have indicated measurable contamination from the primary up to apertures bigger than 18$\times$FWHM; the lack of this contamination was in line with the separation of the primary and with the outcome of our contamination tests.
{We chose the smallest aperture (7 FWHM) for which the difference of flux  between incrementally larger apertures was below 1\%}, indicating that the aperture is capturing most of the intensity in the line spread function.}  

{In Figure \ref{LC_AB} we show the resulting broad band light curves for LP261-75A and LP261-75B.}


\subsection{Pointing stability}

{Our flux density measurements might be also influence by the pointing stability of the instrument plus telescope system within a given orbit}.  To assess the magnitude of this effect, we measured the central positions  of the spectra  in each exposure in all six orbits, {fitting a Gaussian function to the spatial direction in six different columns}. We concluded that the pointing {jitter and/or drift} of HST was about 1/10 of a pixel or less {within each of the six orbits}  (much smaller than the aperture size we adopted). Therefore, the pointing drift or jitter does not affect {significantly} our measurements. 

\subsection{Ramp correction\label{ramp_correction}}

The most prominent source of systematics on WFC3/IR data is the ``ramp effect``, which consists of an approximately exponential-shape signal that increases with time during every orbit.
{The correction described in \citet{Zhou2017} models the charge trapping process of WFC3 {infrared} detector and calculates a ramp effect systematic light curve, which is used to recover the intrinsic  signal.}
{Each photometric data point was corrected for the ramp effect considering the entire dataset (all data points in every orbit) using the physically-motivated detector charge trap model developed by \cite{Zhou2017}. This approach models the number of charges trapped in a given pixel and their delayed release, allowing for the correction of every orbit (including the first), reducing the ramp effect’s amplitude by at least an order of magnitude. After this correction, the dominant source of noise in our data is the photon noise.}

\subsection{Sky background }

We tested the possibility that time-varying scattered light would affect the light curve we obtained for LP261-75B. To rule out that option, we measured the values of the sky after sky background substraction, and ramp correction, in four different {regions} of 256$\times$30 pixels in the detector. The chosen {regions} span from pixels y~=~87 to 117 {(region 1)}, y~=~117 to 147 {(region 2)}, and pixels y~=~162 to 192 {(region 3)} and y~=~192 to 222 {(region 4)}, avoiding the areas between pixels y~=~0 to 87, and pixels y~=~147 to 162, where the spectra of LP261A and LP261-75B were located. 

We calculated the {Kendall's $\tau$ coefficient} to investigate a potential correlation   between the variations of the sky background after ramp correction and sky substraction, and the ramp-corrected white light curve of LP261-75B  for the four selected regions. We obtained Kendall coefficients close to 0 {(region 1: $\tau$ = -0.15, significance:    0.08;  region 2: $\tau$ = -0.13, significance = 0.12; region 3: $\tau$ = -0.16, significance: 0.056; region 4: $\tau$ = 0.08, significance = 0.37)}, indicating no correlation. {Therefore, we conclude that slight variations in the sky background do not influence measureably the light curve of our target.}

\subsection{Correlation between LP261-75A and B light curves}

Finally, we searched for correlations between normalized the light curves of LP261-75A and LP261-75B after their respective ramp corrections, performed as indicated in Section  \ref{ramp_correction}. We calculated  Kendall correlation coefficients between their white light curves, obtaining a {Kendall's $\tau$ coefficient}  close to 0 {($\tau$ = 0.03, significance: 0.73)}, indicating no correlation between LP261-75A and LP261-75B light curves. {Therefore, we conclude that any potential variations in the measured brightness of LP261-75A do not influence the lightcurve we measured for LP261-75B, i.e., the variations seen in LP261-75B are intrinsic.}

{\section{Origin of the spectro-photometric variability of LP216B}\label{origin}}

In this Section {we evaluate} the possible causes that would  explain the {spectro-photometric variability} found for a mid-L dwarf as LP261-75B:

\begin{itemize}

\item {Binarity: Our HST wavelength calibration (direct) images do not resolve any companion to LP261-75B itself at a separation higher than $\sim$1.25 AU. In case LP261-75B is an unresolved binary system, {we would expect it to be overluminous in a color-magnitude diagram in comparison to its counterparts of similar spectral type. {For LP261-75B, its absolute magnitude ($\mathrm{M_J}$) in $J$-band is 14.6$\pm$0.09, consistent with the absolute magnitude expected for a L7 dwarf (the spectro-photometric $\mathrm{M_J}$ for a L7 is 14.3$\pm$0.4, using \cite{Dupuy_Liu2012} spectro-photometric relation.) Therefore, we conclude, that binarity is most probably not the cause of the variability in the light curve of LP261-75B.}}}


{\item Magnetic spots: In L-dwarfs, the magnetic Reynolds number, a parameter that describes how efficient a gas interacts with a magnetic field,  is too small to support the formation of magnetic spots on L-dwarf atmospheres \citep{Gelino2002}. In addition, \cite{Miles_Paez2017} concluded that chromospheric activity and photometric variability are not correlated, specially for objects with spectral types later than L3.5. {In the case of LP261-75B, no evidence of H-$\alpha$ emission  has been found \citep{Kirkpatrick_2000}, indicating a lack of magnetic activity}. Therefore, we do not expect that the photometric variability found for LP261-75B is caused by magnetic spots.}

{\item Heterogeneous cloud coverage: \cite{Ackerman_Marley2001} proposed that heterogeneous clouds similar to those found in the atmospheres of giant planets of our Solar System may be present in the atmospheres of brown dwarfs as well. In fact, photometric variability has been observed for brown dwarfs from mid-L to the late-T spectral types \citep{Artigau2009, Radigan2012, Apai2013, Buenzli2015, Metchev2015,Yang2015,Lew2016}. Inhomogeneous cloud coverage has been {found} to be the {most plausible} cause of photometric variability in L/T transition brown dwarfs \citep{Radigan2012,Apai2013}. For the mid-L dwarfs, \cite{Yang2015} showed that their photometric variability could be due to high altitude haze clouds above condensate clouds. {Therefore, we conclude that the most plausible cause of spectro-photometric variability is the existence of heterogeneous cloud coverage in the atmosphere of LP261-75B.}} 

\end{itemize}

\section{Results}\label{Results}

\subsection{Rotational Period}\label{Rotational_period}

\subsubsection{Lomb-Scargle periodogram}\label{LS}

 {We {determined} the rotational period through a periodogram   using the {IDL\textregistered\textit{periodogram.pro}} function. This routine employs the  method described by \cite{Horne_Baliunas1986}, based on the  Lomb-Scargle periodogram \citep{Lomb1976, Scargle1982}, to calculate a periodogram within user-set frequency or period limits for a time series of data.
{The data do not need to be equally spaced. {The power obtained is normalized to the variance of the total data.} The periodigram derived from the LP261-75B light curve (as shown in Fig. \ref{LC_AB}) is illustrated in Figure \ref{periodogram}. We retrieved a principal period of 4.87$\pm$0.25~h, and a shorter period of significance of 1.19$\pm$0.01~h, with uncertainties as computed by \cite{Horne_Baliunas1986} (ibid, c.f. their Eq. 14). The principal period is close to 3x the HST orbit period of 1.59~h), and the shorter periods are close (but not exactly equal) to harmonics of the principal period. } 

\begin{figure}
\centering
\includegraphics[width=0.45\textwidth]{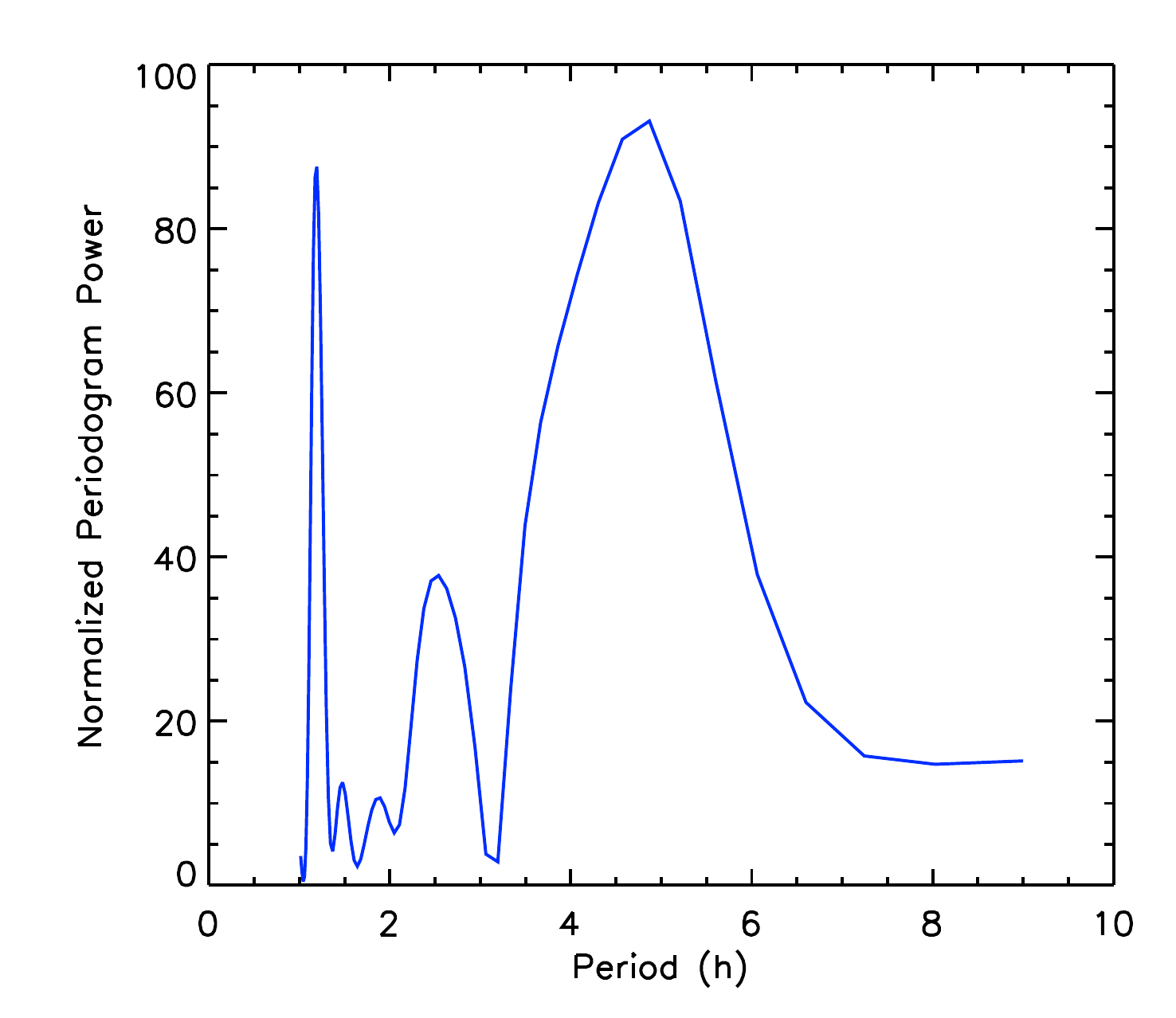}
\caption{\label{periodogram}  Periodogram of the white light curve of LP261-75B. The maximum corresponds to the period of the light curve, that is 4.87$\pm$0.25~h.}
\end{figure}

\subsubsection{Bayesian generalized Lomb-Scargle periodogram}\label{BGLS}

The Lomb-Scargle periodogram does not take  into account {uncertainties in flux}. The Lomb-Scargle periodiagram fits a sine wave to the data to determine the most probable periods.  However it assumes that the mean of the data and the mean of the sine wave fitted to the data to obtain the most probable periods is the same. In addition,  the Lomb-Scargle periodogram as presented in Section \ref{LS} is expressed in an arbitrary power.  These deficiencies have been addressed by several authors \citep{Ferraz-Mello1981, Cumming1999, Bretthorst2001, Zechmeister2009, Mortier2015}. In addition, \cite{Cowan2017} concluded that the periodogram by \cite{Horne_Baliunas1986} provide biased results in case of temporal gaps in the data. 

\cite{Mortier2015}\footnote{https://www.astro.up.pt/exoearths/tools.html} provided a  Python-based program to calculate the Bayesian Generalized Lomb-Scargle periodogram (BGLS) of time-series based on \cite{Bretthorst2001} and \cite{Zechmeister2009}. It provides the relative probabilities between different peaks of similar power in the LS periodogram, taking into account the uncertainties of the data, possible gaps, and {any possible zero point difference in the data collected at different epochs}. The periodogram is shown in Figure \ref{periodogram_Mortier}. We retrieved a principal period of 4.78$\pm$0.37~h, and other shorter period of significance of 1.19$\pm$0.04~h, with uncertainties as computed as the FWHM of each of the peaks after fitting a gaussian to each of them.

\begin{figure}
\centering
\includegraphics[width=0.45\textwidth]{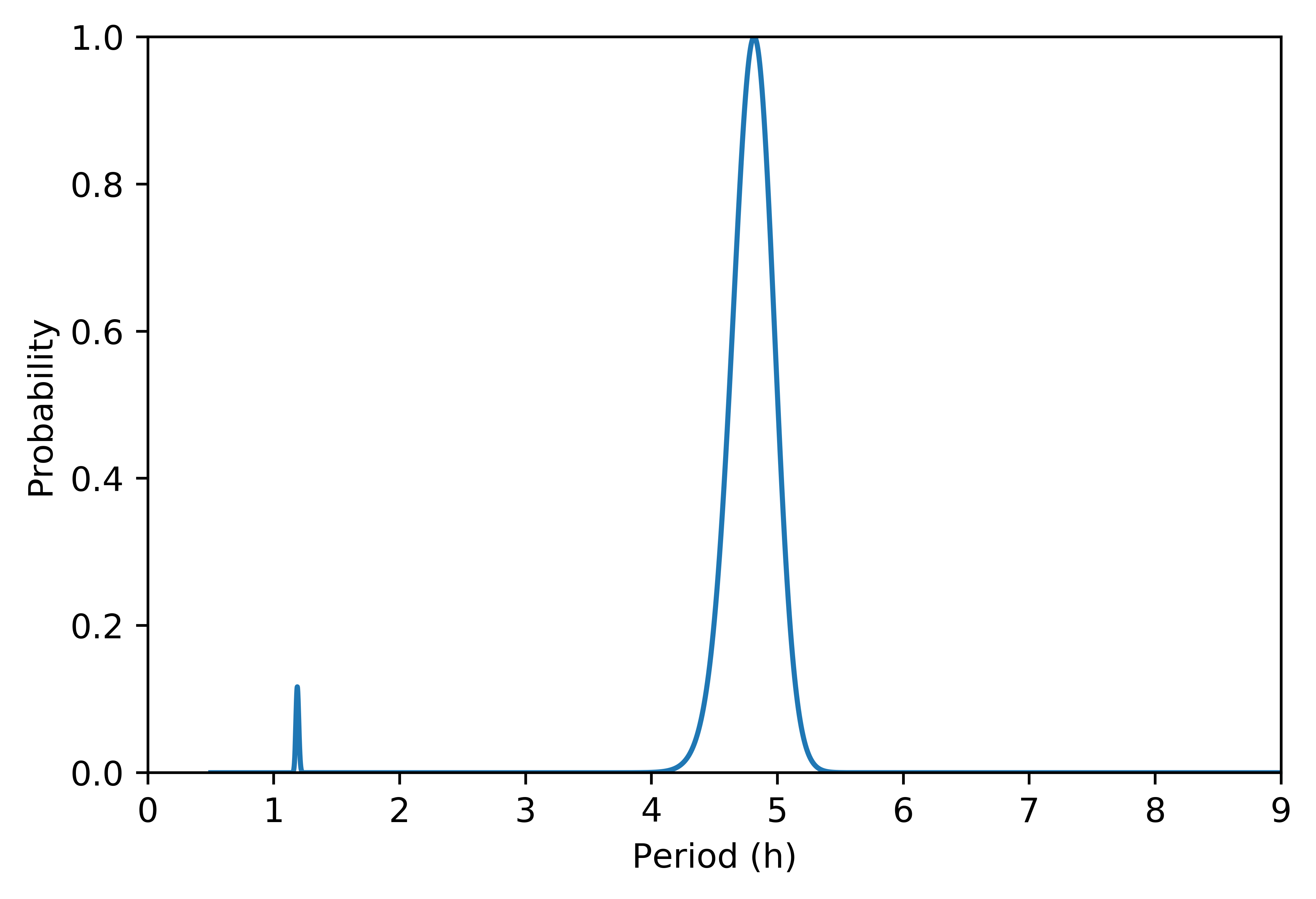}
\caption{\label{periodogram_Mortier}  {Bayesian generalized Lomb-Scargle periodogram of the white light curve of LP261-75B. The y-axis shows the probability that a signal with a specific period is present in the data.  Probabilities are normalized to the probability of the highest peak. The period obtained for the light curve is 4.78$\pm$0.37~h.}}
\end{figure}

\subsubsection{Fourier Function Fit}\label{fourier_fit}

{Recently, \cite{Apai2017} showed that light curve evolution observed in three L/T transition brown dwarfs with near-continuous Spitzer Space Telescope photometry were well-described by planetary-scale waves with periods similar to but slightly different from the likely rotational periods of the objects. This model successfully fitted the light curve evolution as k=1 and k=2 (k - wavenumber) waves. In the following we also explore fitting the LP261-75B data with a Fourier series motivated by the results of Apai et al (2017), without necessarily adopting the same interpretation.}

{We fitted a  1st {to} 5th degree Fourier function fit to the white light curve  using the following expression:}

\begin{equation}
   F(t) = \frac{A_{0}}{2} +  \sum_{n=1}^{1\, to \, 5}\biggl[A_{n} \cos\biggl(\frac{2\pi n}{P}t\biggl)+B_{n}\sin\biggl(\frac{2\pi n }{P}t\biggl)\biggl]
 \end{equation}
 
{We performed  Levenberg-Marquardt least-squares fit to the Fourier function, using the IDL function {mpfitfun.pro \citep{Markwardt2008}}. We chose the 4th Fourier function as the best fit the white light curve, as it gave the lowest reduced $\chi^2$ ($\chi^2_{red}$ = 1.06) and the lowest Bayesian Information Criterion {\citep[BIC]{2007MNRAS.377L..74L}}. {In Table \ref{fourier_fit_chi} we summarize the values of the $\chi^2_{red}$ and the BIC for each of the Fourier functions we fitted to the white light curve.}}
  {For the smallest $\chi^{2}_{red}$ and BIC fit, we obtained a period   of 4.76$\pm$0.03~h, in statistical agreement with the principal period determined from the  two periodogram analysis. {In Table~\ref{fourier_fit_ampl} we show the values of the Fourier components calculated for the 4th order Fourier fit}. In Figure \ref{LC}, top plots, we show the non-folded light curve on the left, and the phase-folded light curve on the right, {using the result of the period found from the 4th order Fourier fit}. After phase-folding the white light curve, {we retrieved a light curve with hints of three peaks}. The residual of the best-fitting Fourier function was smaller than {0.1\% in amplitude, that is of the order of magnitude of the uncertainty level per spectral bin (0.1-0.3 \%)} (see Figure \ref{LC}, bottom plots).} 
  
{Although the principal periods retrieved by the {three methods} are in statistical agreement,  it is important to note that after trying different sets of initial conditions, the Levenberg-Marquardt least-squares fit method   only converged for a specific set of defined initial conditions.}
 
\begin{figure*}[htb!]
\centering
\includegraphics[width=0.8\textwidth]{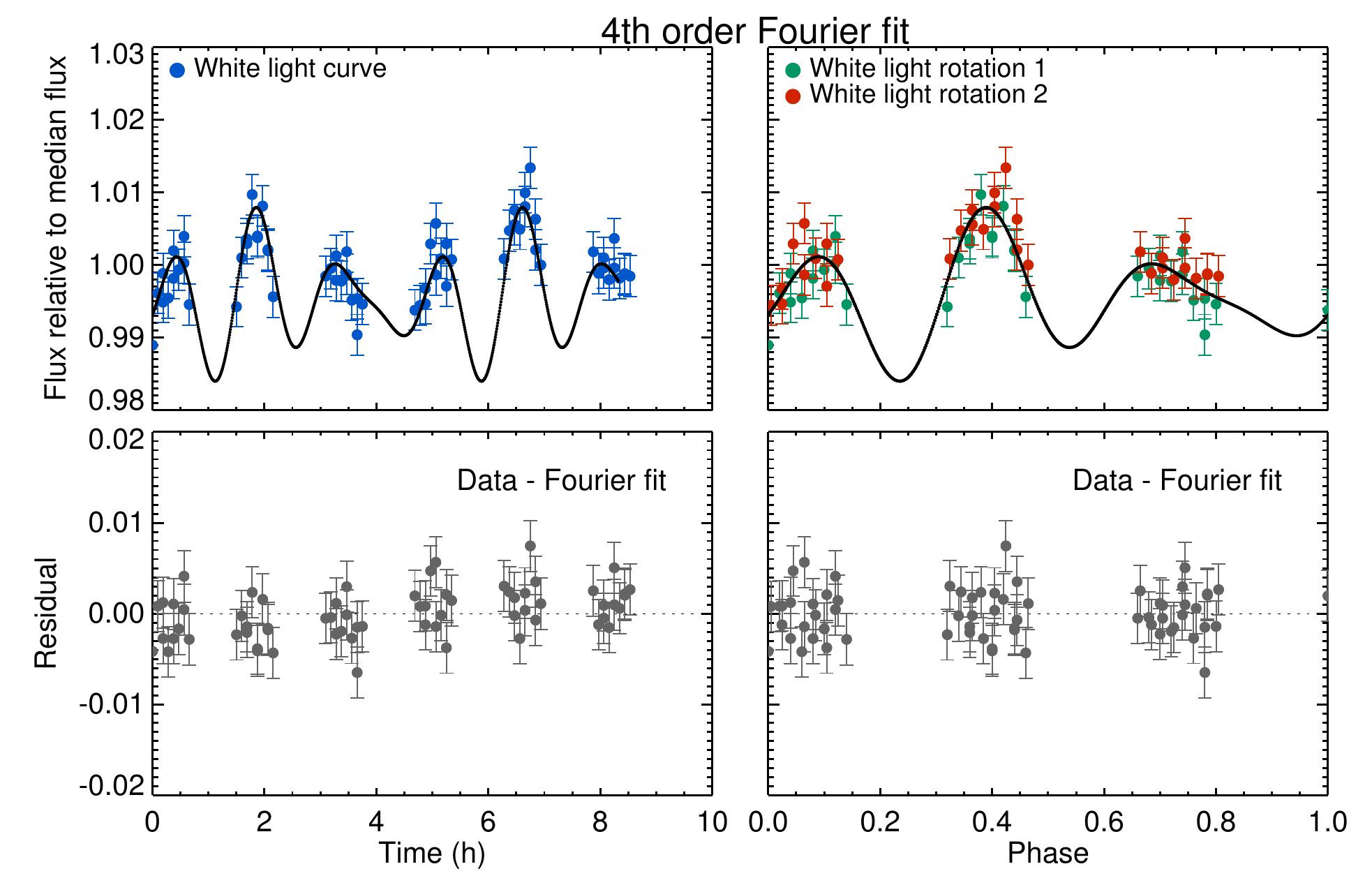}
\caption{\label{LC} Top plots: non-folded light curve of LP261-75B in  white light (1.1--1.69$\mu$m) (left), and folded light curve (right). Overplotted with a black line we show the best fit using a forth order Fourier function.   Bottom plots: residuals after subtracting the best fit with the fourth order Fourier function from the non-folded and folded white light curves.}
\end{figure*} 
 
\begin{table}
	\caption{Reduced $\chi^2$ and BIC (Bayesian Information Criterion) of the Fourier functions fitted to LP261-75B white light curve.}  
	\label{fourier_fit_chi}
	\centering
	\begin{center}
		\begin{tabular}{lll}
        \hline
			\hline 
			
 Order of Fourier Fit & $\chi^2_{red}$ & BIC \\   
 
 \hline
 1st    		   &   	1.67			&	128.8	\\
 2nd			   &	1.69			&	126.6	\\
 3rd			   &	1.24			&	105.8	\\
 4rd			&	1.06				&	101.6	\\
 5th			&	1.10				&	109.9	\\
			\hline			
		\end{tabular}
	\end{center}

\end{table}


\begin{table}
	\caption{Amplitudes for the 4th order Fourier fit.}  
	\label{fourier_fit_ampl}
	\centering
	\begin{center}
		\begin{tabular}{ll}
        \hline
			\hline 
			
 Fourier component &  Value \\   
 
 \hline
  P (h)		 	& 	4.756$\pm$0.030		\\
  $A_{0}$      &   1.991$\pm$0.002				\\
  $A_{1}$	    &	(-1.722$\pm$1.522).$10^{-3}$		\\
  $B_{1}$	  &		(-2.821$\pm$13.653).$10^{-4}$			\\
  $A_{2}$		&	(5.373$\pm$9.786).$10^{-4}$	\\
  $B_{2}$		&	(-1.069$\pm$1.085).$10^{-3}$		\\
  $A_{3}$		&	(2.611$\pm$0.981).$10^{-3}$			\\
  $B_{3}$		 &	(6.372$\pm$1.256).$10^{-3}$			\\
  $A_{4}$		 &	(-3.632$\pm$1.049).$10^{-3}$			\\
  $B_{4}$	 	&	 (-0.200$\pm$1.052).$10^{-3}$		\\
			\hline			
		\end{tabular}
	\end{center}

\end{table}

\subsubsection{Robust estimation of periods and uncertainties}\label{MC}

Due to target visibility interruptions, brown dwarfs{our data do not sample, or closely flank, the troughs} of the phase-folded light curve (Fig. \ref{LC}, top right panel).  We thus present  the subsequent analysis to robustly estimate the significant main periods and their uncertainties using different Monte Carlo simulations:

\begin{enumerate}

  \item Regular Monte Carlo simulation: we generate 1000 synthetic light curves slightly different in flux values from the observed light curve by redefining each data point using a gaussian random number generator. The mean of the gaussian is the measured flux of the original light curve, and the standard deviation is the photometric uncertainty of each point. We plotted  a combined periodogram, obtained as indicated in Section \ref{LS}, of all light curves and overplotted the area between 25\% and 75\% percentiles in light blue (see Figure \ref{periodogram_MC}), and the 50\% percentile in a dark blue line. We obtained a main period of {4.78$\pm$0.95~h} with a {5.7-$\sigma$} detection, and a secondary period of {1.19$\pm$0.06~h}. {The uncertainty on the periods is the FWHM of every of the 50\% percentile peak}. In the right side plot in Figure \ref{periodogram_MC} we show the distribution of the main period obtained for each of the synthetic light curves created in the Monte Carlo simulation. The solid line represent the mean of the distribution, and the dashed lines represent the 1-$\sigma$ of the distribution. 
  
{The reason to employ the regular L-S periodogram presented in Section \ref{LS} instead of the BGLS periodogram presented in Section \ref{BGLS}, is that the BGLS normalizes the probabilities of each of the peaks to the maximum  peak found in each periodogram for each synthetic light curve. Thus, we are not able to compare the periodograms for different light curves, as the normalization factor changes for each periodogram. The power of the peaks in the regular L-S periodogram are normalized to the variance of the data, that should be quite similar for the synthetic light curves generated in the Monte Carlo simulation. Thus, we can compare the different periodograms for the different synthetic light curves.}

\begin{figure}
\centering
\includegraphics[width=0.45\textwidth]{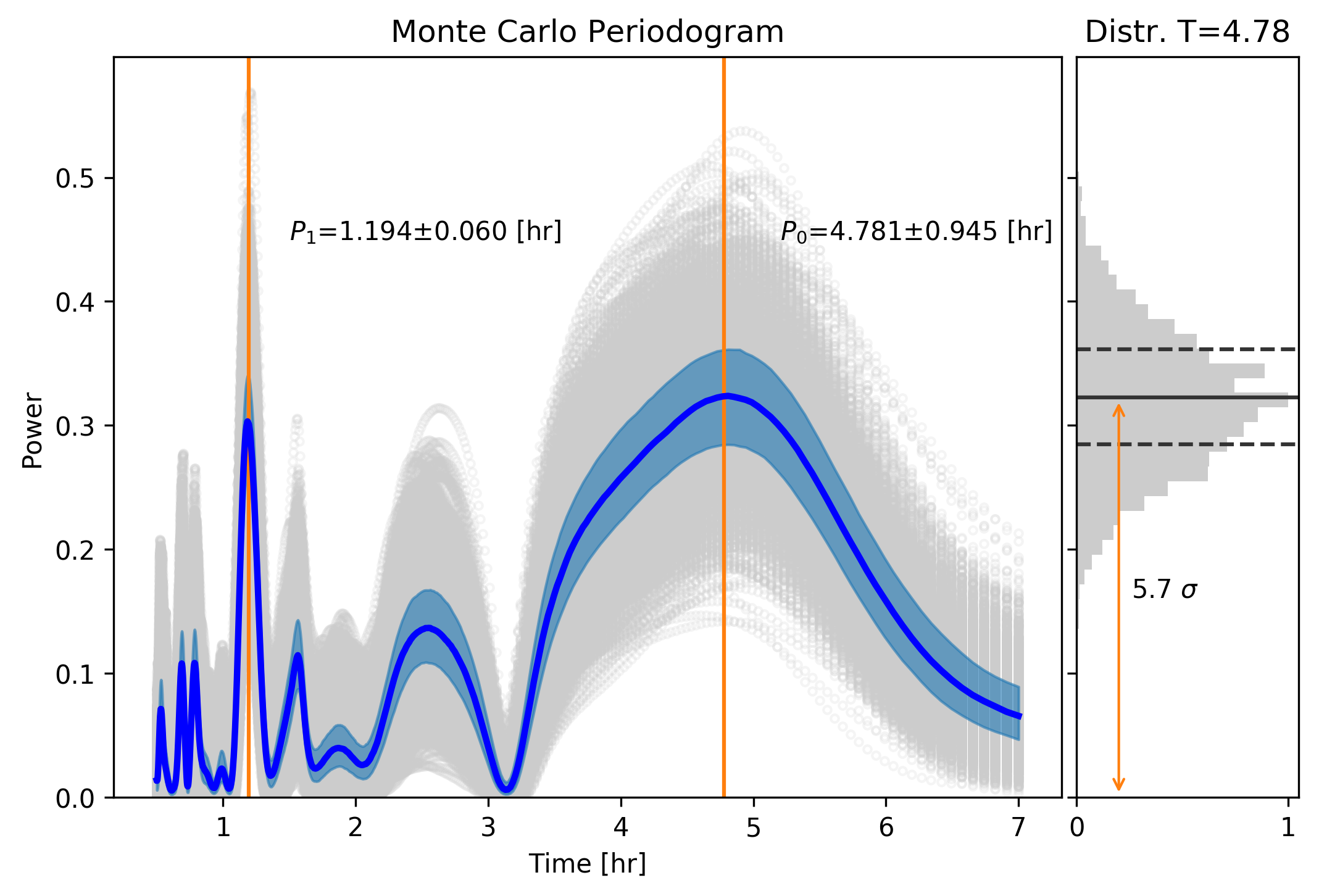}
\caption{\label{periodogram_MC}  Combined periodogram of all the 1000 synthetic white light curves originated in the Regular Monte Carlo simulation. We represent the maximum of the main and secondary peaks with a orange vertical line, that correspond to the two the main and secondary periods found in the Monte Carlo simulation. We overplotted the area between 25\% and 75\% percentiles in light blue, and the 50\% percentile in a dark blue line. In the right side plot we show {the normalized distribution in power of the main period obtained for each of the synthetic light curves} created in the Monte Carlo simulation. The solid line represent the mean of the distribution, and the dashed lines represent the {1-$\sigma$} of the distribution. }
\end{figure}

\item Prayer Bead  Monte Carlo \citep{Moutou2004, Gillon2007}:  this method use the scatter of the residuals to the best model fit, as we obtained in Section \ref{fourier_fit}. We created 1000 new synthetic light curves adding the residuals shifted in time by a different random amount and added to the best model light curve. We plotted a combined periodogram, obtained as indicated in Section \ref{LS}, for the different light curves,  and overplotted the 25\%, 50\% and 75\% percentiles using the same color codes as in Figure \ref{periodogram_PrayerBead}.  We obtained a main period of {4.76$\pm$0.93~h} with a {8.4-$\sigma$} detection, and a secondary period of {1.20$\pm$0.06~h}. 

\item Bootstrap Monte Carlo: this method is similar to the Prayer Bead Monte Carlo, with the difference that the residuals are randomly permuted and added to the best model light curve, as obtained in \ref{fourier_fit}, to create each of the new 1000 synthetic  light curves. We plotted a combined periodogram, obtained as indicated in \ref{LS}, for all light curves and overplotted the 25\%, 50\% and 75\% percentiles (see Figure \ref{bootstrap_systematics}).  We obtained a main period of {4.74$\pm$0.92~h} with  a {6.6-$\sigma$} detection, and a secondary period of {1.19$\pm$0.06~h}.

\end{enumerate}

\begin{figure}
\centering
\includegraphics[width=0.45\textwidth]{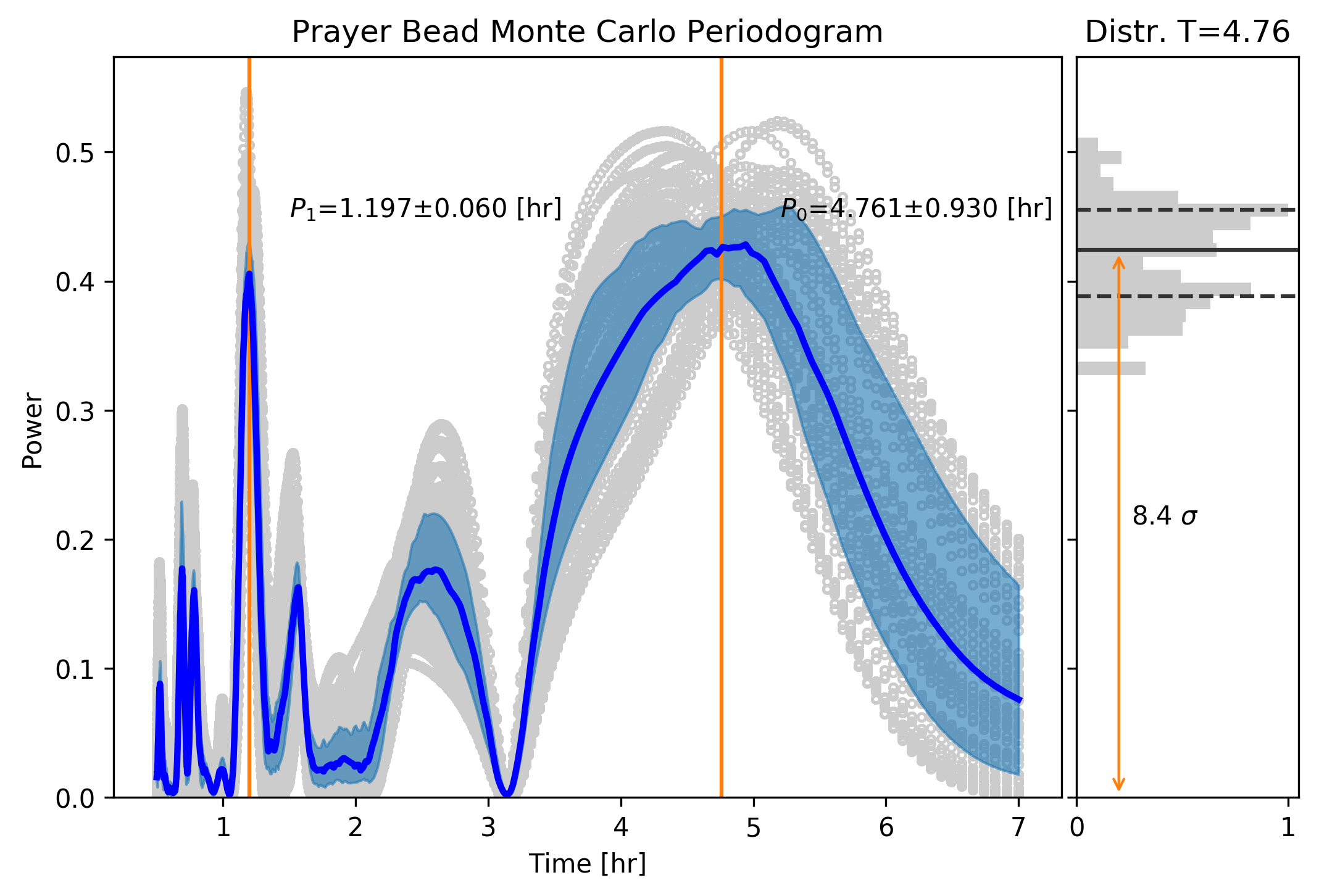}
\caption{\label{periodogram_PrayerBead} Combined periodogram of all the 1000 synthetic white light curves originated in the Prayer Bead Monte Carlo simulation. The color code is the same as in Figure \ref{periodogram_MC}.}
\end{figure}

\begin{figure}
\centering
\includegraphics[width=0.45\textwidth]{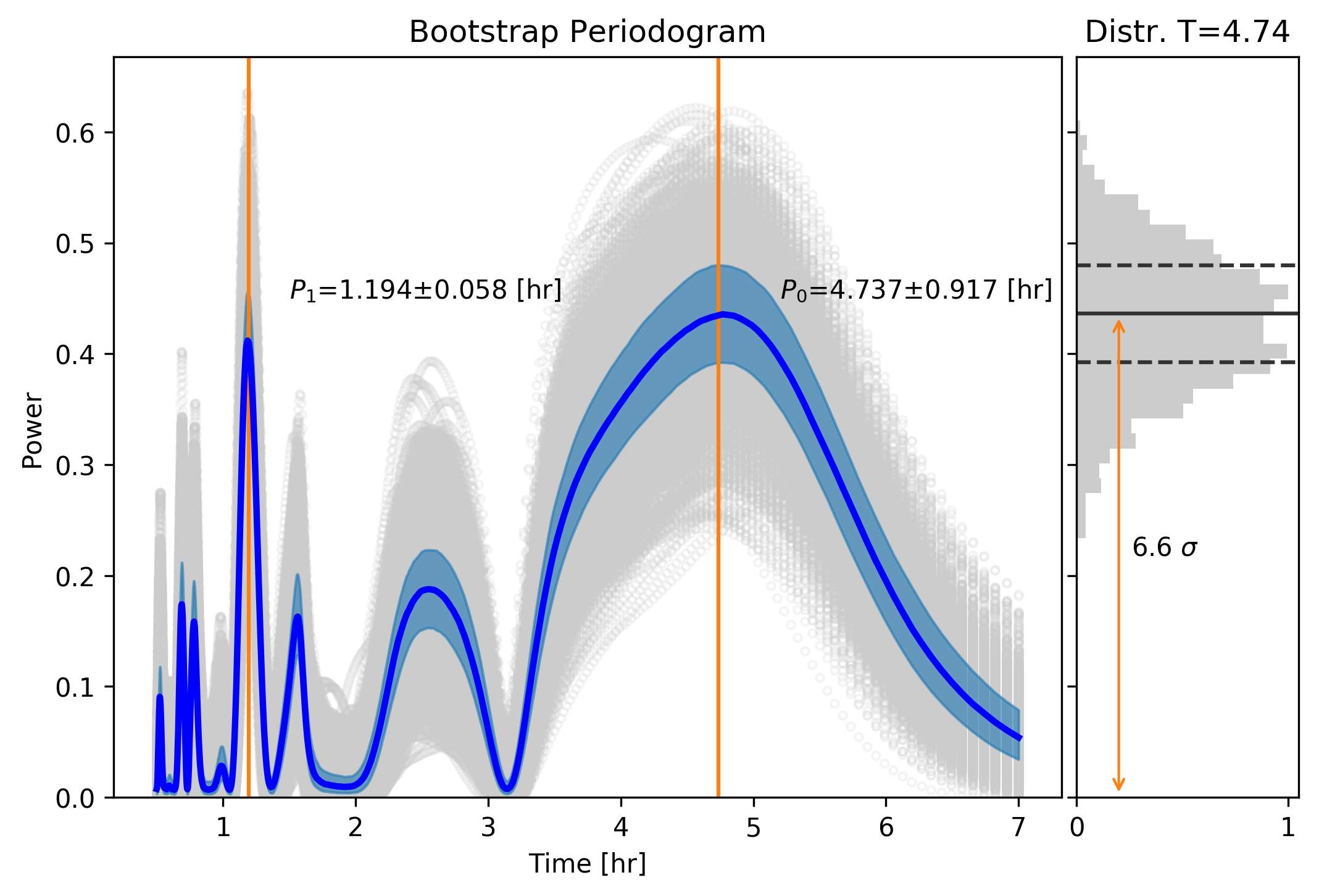}
\caption{\label{bootstrap_systematics} Combined periodogram of all the 1000 synthetic white light curves originated in the Boostrap Monte Carlo simulation. The color code is the same as in Figure \ref{periodogram_MC}.}
\end{figure}

\subsection{Spectral variability \label{spectral_variability}}

We  obtained in total  66 spectra of  LP261-75B during  six HST orbits. 
We explored the  amplitude of the rotational modulations as a function of wavelength by comparing the average of the six spectra with the highest flux and that of the  six with lowest flux (Figure \ref{max_min_spec}, Top Figure).  The middle panel of the Figure \ref{max_min_spec} shows the ratio of the averaged maximum to minimum spectra, i.e., the relative amplitude. The bottom panel shows the result after subtracting a  linear fit found for the ratio between the spectra with the maximum and minimum flux of LP261-75B. {We found a minimum rotational modulation amplitude of approximately 2.41$\pm$0.14\% in the average over the entire wavelength range} and conclude that the variations are gray, i.e., there is no significant wavelength-dependence in the amplitude variations. {The amplitudes variations inside the $\mathrm{H_{2}O}$ band (1.35--1.43~$\mu$m)  and  outside the  $\mathrm{H_{2}O}$ band  are consistent {(2.80 $\pm$ 1.02\%)}}.

\begin{figure}
\centering
\includegraphics[width=0.45\textwidth]{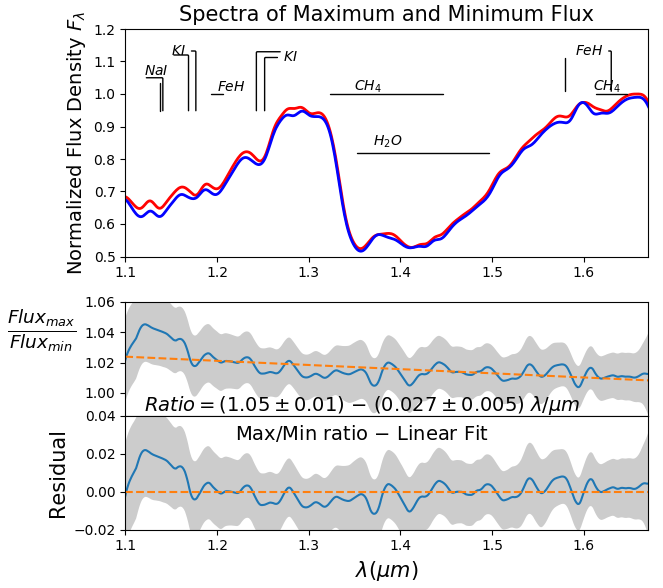}
\caption{\label{max_min_spec} Top panel: average of the six maximum (red color) and the six minimum (blue color) spectra of the 66 spectra taken during six HST orbits for LP261-75B.  Middle panel:  ratio between the six spectra with the maximum flux and six spectra with the minimum flux of LP261-75B, with the best linear fit to it. Bottom panel: residuals of the ratio between the six maximum and six minimum spectra of LP261-75B after subtracting the best linear fit. }
\end{figure}

\section{Discussion \label{Discussion}}

\subsection{Rotational Period}


All methods presented in Section \ref{Rotational_period} to determine the rotational period of LP261-75B light curve  are in statistical agreement, {suggesting} a main rotational period  of $\sim$4.8~h. {Nevertheless, it is important to note that our data do not sample the troughs of the light curve.} In particular,  there are equally-spaced gaps in the light curve that result from the HST-orbit target-occultation period.  This presents a challenge in attempting to reliably determine the LP261-75B rotation period. Contiguous observations over a longer temporal duration (not available from this data set) would more reliably constrain against the possibility of a spurious period determination. Therefore, we adopt {as an indicative period}, the one obtained in the regular Monte Carlo simulation, (4.78$\pm$0.95~h), with its  more conservative uncertainty. 

{In addition, based on the fact that many -but not all- continuous brown dwarf lightcurves show an underlying fundamental period \citep{Metchev2015, Apai2017}, we could expect that the changes found in the lightcurve of LP261-75B are at least quasi-periodic. In facet, as the lightcurves are interpreted as rotational modulations in the literature (see Section \ref{origin}), this means that they will be periodic or quasi-periodic.}

{\subsection{Spectral Modulations and Cloud Structure}}

 In the spectra of ultracool atmospheres, light of different wavelengths probe  different pressure levels. Therefore, spectroscopic rotational modulations inform upon the surface brightness distribution of the atmosphere as functions of longitude and pressure. Spectral contribution functions can be calculated to determine which pressure ranges are probed by which wavelengths. Then, \citet{Yang2016}  used state-of-the-art radiative transfer models to evaluate changes in the emerging spectrum in response to changing temperatures in discrete pressure levels to determine the contribution functions. This approach is imperfect, as it does not include readjustment of the cloud cover and atmospheric dynamics in response to the temperature change introduced, nevertheless it can provide a useful guide for the pressure levels probed for different atmospheres at different wavelengths. 
 
That work (their Figure 16) predicts that for \lpobj the $J$-band will probe the highest pressure level in the atmosphere, with 80\% of the $J$-band emission emerging from pressures 7~bar and lower. The other continuum band in our study, the $H$-band, probes somewhat shallower depths (pressures 6~bar and lower), while the  $\mathrm{CH_{4}-H_{2}O}$ (1.62--1.69~$\mu$m) molecular band traces the atmosphere 5--6~bar and below. The highest opacity in our wavelength range is in the $\mathrm{H_{2}O}$ (1.35--1.43~$\mu$m) band, which probes the atmosphere at pressures $\sim$4~bar and lower. 


Our measurements allow us to explore atmospheric surface brightness variation {\em differences} between the 0--4 bar and 0--7 bar pressure ranges. Our time-resolved spectroscopy found  nearly flat relative amplitude variations in \lpobj.

As shown in Figure~\ref{max_min_spec}, the observed relative amplitude variations for LP261-75B over the 1.1--1.65~$\mu$m  wavelength range have only a {slightly descending slope} ($-0.027\pm0.005$) $\mu m^{-1}$.
The wavelength-dependence of the relative amplitude observed in \lpobj resembles those reported for two L5 high-gravity brown dwarfs (\object{2MASS J15074769--1627386} and \object{2MASS J18212815+1414010}) in \citet{Yang2015} and the L6.5-type intermediate surface gravity, extremely red brown dwarf \object{WISEP J004701.06+680352.1} (hereafter W0047). All four L-type objects show nearly linear wavelength dependence in the relative amplitude of the rotational modulations, with very similar relative changes in the $J$-band and $H$-band continuum and the 1.4$\mu$m water bands. Only W0047 -- the highest-amplitude and also reddest object in the sample -- shows evidence for a somewhat lower modulation amplitude in the water band.

This pattern stands in stark contrast with the sample of L/T transition dwarfs observed with HST: the T2 dwarfs (\object{2MASS J01365662+0933473} and \object{2MASS J21392676+0220226} in \citealt{Apai2013}; and \object{
2MASS J10491891-5319100} or Luhman 16B in \citealt{Buenzli2015}) show greatly reduced modulations in the water band with respect to the modulations observed in the $J$-band and $H$-band continuum bands. 

The amplitudes of the relative modulations argues for a prominent difference between the pressure levels where modulations are introduced, likely through cloud heterogeneities. The fact that there is only small (often undetectable) relative amplitude difference between the water and continuum in L5 dwarfs, argues for the modulation introduced at pressure levels {\em below} (i.e., higher in the atmosphere) than probed in the water band \citep{Yang2015}. This fact, in combination with the observation that the modulations are seen throughout the broad wavelength range of the G141 grism without strong spectral features, led \citet{Yang2015} to argue for the presence of high-altitude, heterogeneous clouds or haze particles in the atmospheres of mid-L-dwarfs.

Our study of \lpobj provides the fourth mid-L-dwarf with observed spectral modulations across this wavelength range. The fact that \lpobj shows relative amplitude modulations very similar to the other three mid-L dwarfs demonstrates that particle cover high in the upper atmosphere is a general characteristic of mid-L dwarfs, as was hypothesized by \citet{Yang2015} on the basis of the first two objects. However, there are also important differences: W0047 shows evidence of weak amplitude reduction in the water absorption band and a strong wavelength-dependent amplitude slope, while the three other mid-L-dwarfs (now including \lpobj) do not. These differences lends further support to the connection between the unusual (extremely red) color of W0047 and its unusual modulations (high-amplitude and with a strong slope), as proposed by \citet{Lew2016}. 

High-altitude small dust grain populations have been proposed previously to explain the red colors and spectra of mid-L-type brown dwarfs.  \citet{Marocco2014}  studied the (time-averaged) spectrum of the extremely red L7pec-type brown dwarf \object{ULAS 222711−004547} and found that its peculiar red color and spectra can be explained by applying a strong reddening slope to an otherwise typical L7-type field brown dwarf spectrum.  \citealt{Hiranaka2016} has extended this study  and presented a systematic comparison of 26 red field brown dwarfs and 26 brown dwarfs with low gravity spectral indicators (spectral types L0-L7). Using a Markov Chain Monte Carlo-based analysis they successfully fitted the objects with combinations of spectral standards and haze (modeled as Mie-scattering particles with smooth power-law particle size distributions). Their fits required grains with $<0.5~\mu m$ diameters.  The analysis of the posterior probability distributions of the haze parameters demonstrated a difference in particle size distributions between the field brown dwarfs and the low gravity brown dwarfs, suggesting a gravity dependence in the processes that form or regulate the hazes.

Our findings provide a strong, independent support to this picture: The time-resolved observations allow us to compare opacity variations within {\em individual} objects, a powerful way to isolate the effects of atmospheric opacity changes from other effects (surface gravity, abundance, vertical mixing) that may be important when comparing different objects to each other.  Our observations directly probe the opacity differences without the complicating effects of changes in bulk parameters and also help determine the pressure levels where the small particles must be present.

{\subsection{Interpretation of LP621-75B light curve}}


Previous mapping efforts of brown dwarf atmospheres using observed light curves showed that a number of cloud heterogeneities, hereafter ``spot'',  with different sizes and at different locations on the disk can reproduce the observed light curve modulations \citep{Apai2013,Karalidi2015}. However, the light curve of LP261-75B poses a challenge for the standard mapping techniques, since it is physically unlikely that isolated spots cause the posited three-peaked
light curve. 

In fact, the lightcurve appears to be, on the face of it, impossible to explain with a rotating brown dwarf with brightness markings.  To understand this, we follow \cite{Cowan2017} in adopting the formalism of Fourier analysis, where $n=1$ denotes the fundamental mode (the brown dwarf's rotational period) and the harmonics are denoted by $n=2,3,$ etc. {The amplitudes of each of the Fourier frequencies} are: $\tilde{A_{n}} = \sqrt{A_{n}^{2}+B_{n}^{2}}$ where $A_{n}$ and $B_{n}$ are the amplitudes in Table \ref{fourier_fit_ampl}.  If its rotational period is 4.8~hrs, then the lightcurve of LP261-75B is dominated by the $n=3$ term, namely three peaks per rotation.  We denote the amplitude at this frequency as $\tilde{A_3}$ ; what we report in practice is the amplitude normalized by the time-averaged flux of the brown dwarf, $\tilde{A_0}=A_0/2$. For the current lightcurve, $\tilde{A_1}/\tilde{A_0} \approx 1.7\times 10^{-3}$ and $\tilde{A_3}/\tilde{A_0} \approx 6.9\times 10^{-3}$. 

It is intuitive to imagine that three bright spots equally spaced in longitude could produce the lightcurve of LP261-75B, but that is incorrect: such a map would in fact produce $\tilde{A_3}=0$.  In fact, it has been shown analytically that a rotating brown dwarf {seen} from an equatorial viewpoint  or with a N--S symmetric map cannot produce non-$\tilde{A_3}$ \citep{Cowan_Agol2008,Cowan2013}---the so-called ``odd nullspace''.  

\cite{Cowan2013} showed that the only way to get non-zero odd harmonics in a rotational lightcurve is with a N--S asymmetric map \emph{and} an inclined viewing geometry. The very simplest map that can produce $\tilde{A_3}$ is the $Y_4^3$ spherical harmonic, but one can achieve a similar effect with six spots (three each in the northern and southern hemispheres, offset in longitude).  

{Spot-based maps tend to produce more power in n=1 than in n=3. Therefore, they may not be able to reproduce our triple-humped lightcurve.} The map--lightcurve convolution is a low-pass filter so higher-order harmonics should be strongly suppressed \citep{Cowan_Agol2008}. A corollary of this low-pass filter is that it is difficult to construct a spot-based map that produces significant $\tilde{A_3}$ without producing at least as much $\tilde{A_1}$ \citep{Cowan2017}, which is inconsistent with the measured amplitudes of LP261-75B ($\tilde{A_3}/\tilde{A_1} \approx 4$).   

We instead explore whether one could explain the lightcurve of LP261-75B with a smoothly-varying brightness map.  In particular, spherical harmonic maps have the advantage of inducing sinusoidal brightness variations at a single frequency: a $Y_l^m$ can only produce power at the $n=m$ harmonic.  Following \cite{Cowan2013}, a pure $Y_0^0+C_4^3 Y_4^3$ map with 13\% semi-amplitude could produce $\tilde{A_3}$ of 1\% if the BD is inclined 60$^\circ$ from pole-on. So the lightcurve of LP261-75B can indeed be produced by a positive map, provided it is fairly smooth.  Since the lightcurve morphology has not yet been firmly established, we limit ourselves to this existence proof and leave more detailed mapping exercises for the future. 


{We acknowledge that the  phase-folded light curve  does not sample, or closely flank, the troughs}, and that the inferred period is a multiple of the sampling period. To confirm the period found, continuous data coverage with a longer baseline is needed.

\section{Conclusions}\label{Conclusions}

 We present six consecutive orbit HST/WFC3/G141 time-resolved spectroscopy observations of the \object{LP261-75B} brown dwarf, an L6-type wide companion (380 au projected separation) to an M4.5 star. The key findings of our study are as follows: 

\begin{itemize}

\item We have discovered rotational modulations in the spectrum of \lpobj, the fourth  variable L-dwarf for which time-resolved {spectroscopy} has been obtained. 

\item {We detected a  rotational modulation with an amplitude of {\em at least} 2.41$\pm$0.14\% in the 1.1--1.69~$\mu$m wavelength range.}

\item Based on extensive frequency analysis {{we adopted an {indicative} rotational period} of 4.78$\pm$0.95~h for LP261-75B}. {Due to the incomplete time coverage of the data, and the fact that only the peaks of the light curve are sampled, contiguous observations over a longer temporal duration (not available from this data set) would more reliably constrain against the possibility of a spurious period determination.}

\item On the basis of the phase-folded light curve we report {a possible triple-peaked} light curve. If confirmed, this unusual light curve is most probably generated by longitudinally nearly evenly distributed spots.


\item The ratio of the six maximum and six minimum spectra for LP261-75B is nearly constant across the wavelength range of 1.1--1.69~$\mu$m, with a mild slope (larger amplitudes in the blue) and with no measurable decrease in the water absorption band. This finding contributes to the evidence that mid-L dwarfs share a similar rotational modulation trend and lends further support to the proposed high-altitude clouds or hazes in these objects \citep{Yang2015}.


\end{itemize}

\acknowledgments
{We thank our referee Dr. Derek Homeier for his valuable comments that improved this work}.
Based on observations made with the NASA/ESA Hubble
Space Telescope, obtained at the Space Telescope Institute, which is operated by AURA, Inc., under NASA contract NAS 5-26555, under  GO-14241. This publication makes use of data
products from the Two Micron All Sky Survey, which is a joint
project of the University of Massachusetts and the Infrared
Processing and Analysis Center/California Institute of Technology,
funded by the National Aeronautics and Space
Administration and the National Science Foundation.

\software{IDL\textregistered\textit{periodogram.pro}, mpfitfun.pro \citep{Markwardt2008}}.

\bibliographystyle{apj}
\bibliography{LP621_variability}
\end{document}